\documentclass[twocolumn,aps,pra,superscriptaddress,footinbib,amsmath,amssymb,showpacs]{revtex4-1} %
\usepackage{graphicx}
\usepackage{bm}
\usepackage[colorlinks,	linkcolor=blue,anchorcolor=blue,citecolor=blue,bookmarksnumbered]{hyperref}
\usepackage{float}
\usepackage{amsthm,amsmath,amssymb}
\usepackage{mathrsfs}
\usepackage{xcolor}
\usepackage{titlesec}

\usepackage{babel}

\begin{document}
\title{Routing thermal noise flow and ground-state cooling in an optomechanical
plaquette}
\author{Guang-Zheng Ye}
\address{Fujian Key Laboratory of Quantum Information and Quantum Optics and
Department of Physics, Fuzhou University, Fuzhou 350116, People\textquoteright s
Republic of China}
\author{Tian-Le Yang}
\address{Fujian Key Laboratory of Quantum Information and Quantum Optics and
Department of Physics, Fuzhou University, Fuzhou 350116, People\textquoteright s
Republic of China}
\author{Wan-Jun Su}
\address{Fujian Key Laboratory of Quantum Information and Quantum Optics and
Department of Physics, Fuzhou University, Fuzhou 350116, People\textquoteright s
Republic of China}
\author{Yong Li}
\email{yongli@hainanu.edu.cn}
\address{Center for Theoretical Physics \& School of Physics and Optoelectronic
Engineering, Hainan University, Haikou 570228, China}
\author{Huaizhi Wu}
\email{huaizhi.wu@fzu.edu.cn}
\address{Fujian Key Laboratory of Quantum Information and Quantum Optics and
Department of Physics, Fuzhou University, Fuzhou 350116, People\textquoteright s
Republic of China}
\begin{abstract}
We propose an effective method for cooling two non-degenerate mechanical
resonators by routing thermal noise flow in a four-mode optomechanical
plaquette. The thermal noise flow between the mechanical resonators
can be fully suppressed by addressing the overall loop phase in the
plaquette, irrespective of their thermal temperatures.
We find that optimal mechanical cooling, even down to the ground state,
can be realized in this regime. The thermal noise routing, achieved by dissipation engineering at optomechanical interfaces, provides a valuable and complementary approach to conventional coherent dark-mode control theory. It can be generalized to nonreciprocal control of phonon transport and mechanical cooling, and may find applications in optomechanical networks with complex thermal environments.
\end{abstract}
\maketitle
\textit{Introduction} - Optomechanics, which explores the interaction
of mechanical motion with light, has many applications in fundamental
and applied physics \citep{Aspelmeyer2014,Barzanjeh2022}. The compatibility with a wide range of frequencies enables mechanical motion coupling
to a wide variety of natural or engineered quantum systems in optical
or microwave domain \citep{Kurizki2015,Blais2021,Barzanjeh2022},
among which cavity optomechanical systems are not only an ideal platform
for studying macroscopic non-classical properties \citep{Teufel2011,Liao2016,Riedinger2018,Hu2019,Kotler2021a,Thomas2021},
and testing the fundamental quantum theory \citep{Bassi2013,Vivoli2016,Marinkovic2018},
but also are promising for quantum information processing \citep{Braunstein2005,Rosenberg_2009,Reed2017,Pfaff2017,Fiaschi2021},
high-precision measurement \citep{Giovannetti2001,LaHaye2004,Giovannetti2004,Zhang2012,Peano2015,Motazedifard2019,Clarke2023},
and frequency-conversion transducer \citep{Xu2016,Malz2018,Lauk2020,Lambert2020}.
In particular, the progress in fabrication techniques makes it possible
to couple multiple mechanical resonators (MRs) to electromagnetic
radiation by embedding nano- or micro-scale MRs in optical cavities
\citep{Chan_2011,Verhagen_2012} or superconducting microwave circuits
\citep{Wollman2015,Pirkkalainen2015,Barzanjeh2019}. This helps to
bring the so-called multimode optomechanical systems to the fore as
candidates for studying collective synchronization, macroscopic entanglement,
and other quantum many-body effects \citep{Heinrich2011,Xuereb2012,Ludwig2013,Xuereb2014a,Xuereb2015,Cernotik2018,Carollo2020}.

The practical applications of optomechanical systems in the quantum regime relies on the capability of controlling the optomechanical
interaction in a coherent way \citep{Dowling_2003}, where cooling
MRs to the motional ground state and suppression of the environmental
thermal noise are the prerequisites for observing and manipulating
quantum mechanical effects \citep{Mancini2003,Vitali2007,Lai2022a,Lai2022b}.
Strategies for ground-state cooling of mechanical modes have been
widely studied \citep{Wilson-Rae2007,Marquardt2007,Delic2020a,Toros2021,Genes2008,Dong2015,Lai2021a,Mancini1998,Cohadon1999,Kleckner_2006,Corbitt2007,Poggio2007,Guo2014,Gu2013,Liu2013a,Liu2015,Naseem2021,Liu2025}
(e.g. by using optical sideband cooling \citep{Wilson-Rae2007,Marquardt2007,Delic2020a,Toros2021},
feedback-aided cooling \citep{Mancini1998,Cohadon1999,Kleckner_2006,Corbitt2007,Poggio2007},
and reservoir engineering \citep{Gu2013,Liu2013a,Liu2015,Naseem2021,Liu2025})
and experimentally realized in different architectures, e.g. optical
cavity \citep{Schliesser2008,DelosRiosSommer2021}, microwave circuit
\citep{Teufel2011a,Massel2012,Damskagg2019} and integrated photonic
crystal \citep{Chan_2011,Guo_2019}. For multimode optomechanical
setup, optomechanical sideband cooling of two or more MRs with degenerate
or near-degenerate frequencies may be inefficient, partially due to
the formation of mechanical dark modes \citep{Lai2018,Ockeloen-Korppi2019,Lai2020a,Huang2022,Xu2022,Lai2022,Wen2022,Liu2022a,Cao2025},
where some of the hybrid mechanical modes are decoupled to the cold
optical reservoir. Dark-mode breaking can be implemented by introducing
laser detunings and complex coupling structures to couple the dark
mechanical modes \citep{Lai2020a,Huang2022} or using auxiliary cavity
modes \citep{Liu2022a,Cao2025}, but it is strongly based on the coherent
control of the optomechanical couplings. The dark-mode description
is typically applied for (nearly) degenerate MRs with similar damping
rates. In the case of non-degenerate MRs coupled to thermal baths
at identical temperatures, thermal-phonon-number-dependent effective
damping rates can vary considerably between the MRs, thereby hindering
the preservation of mechanically dark modes. Alternatively, optomechanical
cooling can also be realized by introducing chiral optomechanical
damping \citep{Kim2017} or nonreciprocal phonon transmission \citep{Xu2019},
however, the cooling effect for multiple MRs is still far from the
sideband cooling limit. Moreover, it is unclear when the mechanical
modes are subject to the thermal environments with different temperatures
and how the thermal phonon flow control can benefit the mechanical
cooling, which provides the basis for quantum information processing
with a distributed optomechanical network.

In this paper, we consider an optomechanical plaquette comprising
two MRs and two intermediate cavities, where the MRs have largely
different frequencies. As such, the mean thermal phonon numbers of
the two MRs differ significantly under the same environmental temperature,
and become equal only when the resonators are placed in distinct thermal
environments. For the latter the typical dark-mode description can
still apply. We propose an effective method to cool mechanical motion
in the resolved-sideband regime based on the control of the thermal
noise flow \citep{Barzanjeh2018}. We show that, by addressing the
overall ``plaquette phase'' introduced by laser drivings, the noise
flow between the two MRs can be completely suppressed when the coherent
optomechanical coupling strengths and cavity decay rates satisfy a
specific impedance matching condition. As a result, the ground-state
cooling can be achieved despite differences in mean thermal phonon
numbers, and can approach the dual-cavity cooling limit - the optimal
cooling performance for a MR coupled to two cavity modes. The proposed
method is conceptually distinct from dark-mode engineering techniques;
whereas the latter concerns the manipulation of coherent optomechanical
coupling, our approach specifically targets the control and utilization
of dissipative coupling channels. Our approach has remarkable flexibility
in thermal noise control, and can be applied to nonreciprocal noise
flow and manipulation of phonon transport.

\begin{figure}[H]
\centering{}\includegraphics[width=0.9\columnwidth]{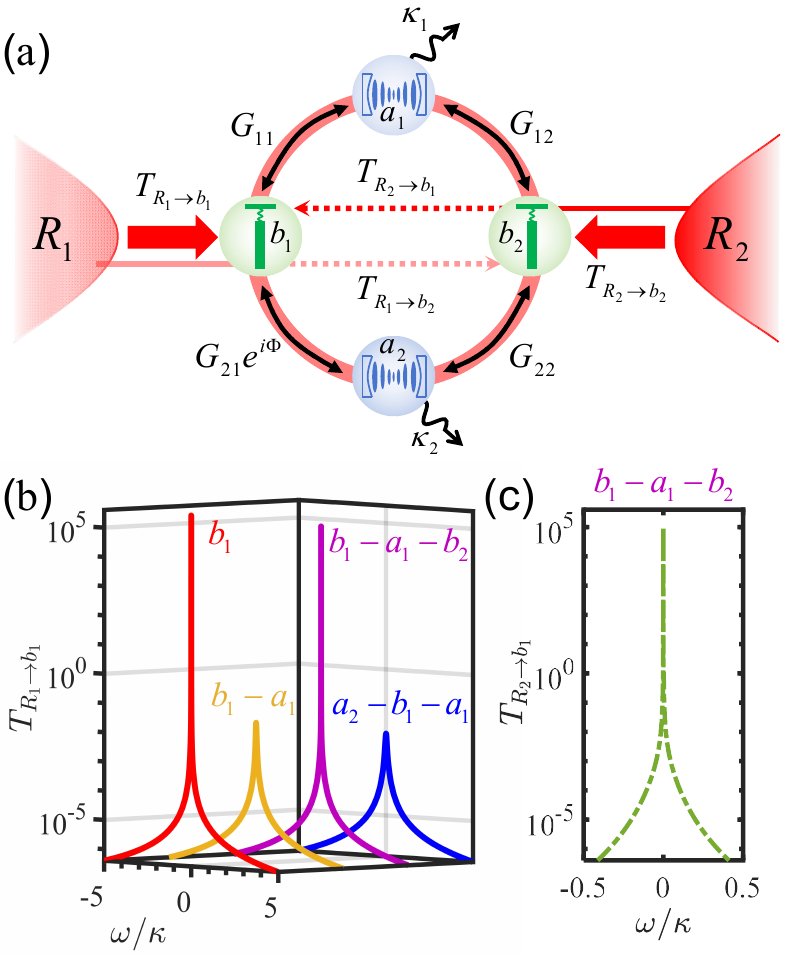}\caption{\label{fig:model}(a) Schematic of a four-mode optomechanical system
consisting of two optical modes ($a_{1}$ and $a_{2}$) and two mechanical
modes ($b_{1}$ and $b_{2}$) with their respective reservoirs ($R_{1}$
and $R_{2}$). $G_{jk}$ ($j,k$ $\in$ $\{1,2\}$) are field-enhanced
optomechanical coupling strengths, and $\Phi$ is the overall phase
induced by the phase-correlated driving lasers. The thermal noise
flow from the reservoir $R_{l}$ to the mechanical mode $b_{k}$ is
described by the scattering amplitude $T_{R_{l}\rightarrow b_{k}}$
($l,k$ $\in$ $\{1,2\}$). (b) $T_{R_{1}\rightarrow b_{1}}(\omega)$
as functions of $\omega/\kappa$ for two-mode and three-mode series
setup: $b_{1}$ (red solid), $b_{1}-a_{1}$ (yellow solid), $b_{1}-a_{1}-b_{2}$
(purple solid), and $a_{2}-b_{1}-a_{1}$ (blue solid). (c) $T_{R_{2}\rightarrow b_{1}}$
versus $\omega/\kappa$ for the $b_{1}-a_{1}-b_{2}$ model, see details
in the main text. Parameters in units of $\kappa=2\pi\times1$ MHz
are $\Delta_{1}=\Delta_{2}=0$, $G_{jk}/\kappa$ $=$ $G/\kappa$
$=$ $0.1$, $\gamma_{1(2)}/\kappa$ $=$ $10^{-5}$, and $\kappa_{1(2)}/\kappa=1$.}
\end{figure}

\textit{Model} - We consider an optomechanical plaquette {[}cf. Fig.
\ref{fig:model}(a){]}, comprising two non-degenerate MRs (with motional
frequencies $\omega_{b,1}$ and $\omega_{b,2}$, and intrinsic damping
rates $\gamma_{1}$ and $\gamma_{2}$), coupled via two cavity modes
(with frequencies $\omega_{a,1}$ and $\omega_{a,2}$, and decay rates
$\kappa_{1}$ and $\kappa_{2}$, respectively). The cavities are driven
at the frequencies $\omega_{a,j}-\omega_{b,k}+\Delta_{k}$ (\{$j,k$\}=1,2)
close to the red mechanical sidebands (with $\Delta_{k}$ $\ll$ $\omega_{b,k}$,
$|\omega_{b,1}-\omega_{b,2}|$). Under the resolved sideband regime
($\omega_{b,k}$ $\gg$ $\{\kappa_{j},\gamma_{k}\}$) and weak coupling
conditions, we can derive the linearized Hamiltonian ($\hbar=1$)
under the rotating-wave approximation (RWA) \citep{Xu2016,Damskagg2019}
\begin{eqnarray}
H_{1} & = & \sum_{k=1,2}\Delta_{k}b_{k}^{\dagger}b_{k}+\sum_{j,k}G_{jk}a_{j}b_{k}^{\dagger}+\text{H.c.},\label{eq:H_1}
\end{eqnarray}
where $a_{j}$ and $a_{j}^{\dagger}$ ($b_{k}$ and $b_{k}^{\dagger}$)
are the annihilation and creation operators of the cavity modes (mechanical
modes). $G_{jk}=g_{jk}\alpha_{jk}$ are the effective optomechanical
coupling strengths, and $\alpha_{jk}$ are the cavity amplitudes under
laser drivings with tunable phases $\phi_{jk}\equiv\text{arg}(\alpha_{jk})$
(see Appendix A).

\textit{Thermal noise flow} - The cavity and mechanical modes are
subject to the zero-temperature bath and the thermal heat bath $R_{k}$
(of temperature $T_{k}$), respectively. For the four-mode optomechanical
plaquette, the net flow of thermal noise into or out of the MRs can
be defined as the difference of the average occupation number of the
MRs ($\bar{n}_{k}=\langle b_{k}^{\dagger}b_{k}\rangle$) to that in
the thermal equilibrium with its own bath $R_{k}$ {[}$\bar{m}_{k}=(e^{\hbar\omega_{b,k}/k_{B}T_{k}}-1)^{-1}${]},
i.e. $\delta n_{k} := \bar{n}_{k}-\bar{m}_{k}$ \citep{Barzanjeh2018}.
When $\delta n_{k}<0$ ($\delta n_{k}>0$), it means the thermal noise
flow out (into) the MR mode $b_{k}$, giving rise to decrease (increase)
of $\bar{n}_{k}$ and cooling (heating) of the $k$th MR. When the
MR approaches the ground state, one has $\bar{n}_{k}\rightarrow0$
and thus $\delta n_{k}\rightarrow-\bar{m}_{k}$.

As an instructive example of mechanical cooling, we consider first
the standard optomechanical setup (denoted as $a_{1}-b_{1}$ setup)
by setting $G_{12}$ $=$$G_{21}$ $=$ $G_{22}$ $=$ $0$. Solving
the quantum Langevin equations for the system in the frequency domain
(see Appendix B), and eliminating the cavity modes, we obtain
\begin{eqnarray}
\left[\chi_{\mathcal{F}}^{(11)}\right]^{-1}b_{1}[\omega] & = & \mathit{\Gamma}_{1}b_{1,in}[\omega]+\mathcal{B}_{11}\mathcal{K}_{1}a_{1,in}[\omega],
\end{eqnarray}
where $a_{1,in}$ and $b_{1,in}$ are the optical vacuum noise and
the mechanical thermal noise, respectively; the beam-splitter-like
optomechanical interaction ($\sim G_{jk}a_{j}b_{k}^{\dagger}$) introduced
by red-detuned laser driving leads to a reduction of the thermal noise
$b_{1,in}$ flowing into $b_{1}$ by the factor of $\chi_{\mathcal{F}}^{(\alpha)}$=$\mathcal{F}(\mathcal{A}_{\alpha}\mathcal{B}_{\alpha})$
with $\mathcal{F}(x)=\left(1-x\right)^{-1}$ and $\alpha=\{jk\}$,
where $\mathcal{A}_{jk}=-iG_{jk}\chi_{aj}$, $\mathcal{B}_{jk}=-iG_{jk}\chi_{bk}$,
$\mathcal{K}_{j}=\chi_{aj}\sqrt{\kappa_{j}}$, and $\mathit{\Gamma}_{k}=\chi_{bk}\sqrt{\gamma_{k}}$
{[}with $\chi_{aj}$ $=$ $(\kappa_{j}/2-i\omega)^{-1}$ and $\chi_{bk}$
$=$ $(\gamma_{k}/2+i\Delta_{k}-i\omega)^{-1}$ being the optical
and mechanical susceptibility{]}. When the cavity vacuum noise is
neglected, the mechanical occupation number then reads $\bar{n}_{1}$
$\approx$ $\bar{m}_{1}\mathcal{T}_{11}$ (where $\mathcal{T}_{11}$
$=$ $\frac{1}{2\pi}\int T_{R_{1}\rightarrow b_{1}}(\omega)d\omega$
with $T_{R_{1}\rightarrow b_{1}}(\omega)=|\chi_{\mathcal{F}}^{(11)}\mathit{\Gamma}_{1}|^{2}$
is the scattering amplitude from its own heat bath). As shown in Fig.
\ref{fig:model}(b), for the set of parameters $\kappa/2\pi=1$ MHz,
$\gamma_{1}/\kappa$ $=$ $10^{-5}$, $\bar{m}_{1}=10^{3}$, $G_{11}/\kappa$
$=$ $0.1$, and $\Delta_{k}/\kappa=0$ (i.e., the red-sideband condition),
$T_{R_{1}\rightarrow b_{1}}(\omega)$ is strongly suppressed due to
optomechanical coupling (the yellow curve), and the net noise flow
out of the MR1 is $\delta n_{1}$ $\approx$ $\bar{m}_{1}\left(\mathcal{T}_{11}-1\right)$
$\approx$ $-\bar{m}_{1}$, which implies the cavity-assisted ground-state
cooling of mechanical motion \citep{Marquardt2007}.

\textit{Series setup} - We then look into the setup with one MR $b_{1}$
coupled to two cavity modes $a_{1}$ and $a_{2}$ (i.e., the $a_{2}-b_{1}-a_{1}$
setup). By setting $G_{12}$ $=$ $G_{22}$ $=$ $0$ in the Hamiltonian
(\ref{eq:H_1}) and following the same procedure above, we find
\begin{eqnarray}
\left[\chi_{\mathcal{F}}^{(11,21)}\right]^{-1}b_{1} & = & \varGamma_{1}b_{1,in}+\sum_{j=1,2}\mathcal{B}_{j1}\mathcal{K}_{j}a_{j,in},\label{eq:NoiseFlow_aba}
\end{eqnarray}
where $\chi_{\mathcal{F}}^{(11,21)}=\mathcal{F}\left(\sum_{j=1,2}\mathcal{A}_{j1}\mathcal{B}_{j1}\right)$.
In comparison with the $a_{1}-b_{1}$ setup, the MR coupled to two cavity modes allows for a better cooling effect, which is manifested
by the decrease of the susceptibility $\chi_{\mathcal{F}}^{(11,21)}$
and the reduction of the response from $R_{1}$ given by $T_{R_{1}\rightarrow b_{1}}(\omega)$
$=$ $|\chi_{\mathcal{F}}^{(11,21)}\mathit{\Gamma}_{1}|^{2}$. As
shown in Fig. \ref{fig:model}(b), for $G_{21}=G_{11}$, the peak
value ($6.25\times10^{-3}$) of $T_{R_{1}\rightarrow b_{1}}(\omega)$
(the blue curve) is about a quarter of that ($2.5\times10^{-2}$)
for the $a_{1}-b_{1}$ setup, while the full-width-at-half-maximum
almost becomes double. Hence, this setup allows for cooling the MR
$b_{1}$ to a lower temperature, which is referred to as the dual-cavity
cooling limit.

There exists another three-mode scenario, which comprises of two MRs
and a cavity mode (i.e., $b_{1}-a_{1}-b_{2}$). Here, the thermal
flow into MR1 comes no only from its own heat bath $R_{1}$, but also
indirectly from the heat bath $R_{2}$ for MR2. By setting $G_{21}$
$=$ $G_{22}$ $=$ $0$, the coupled equation of the two MRs is given
by
\begin{eqnarray}
\chi_{B}^{-1}\left(\begin{array}{c}
b_{1}\\
b_{2}
\end{array}\right) & = & \left[\begin{array}{cc}
1 & \mathcal{A}_{12}\mathcal{B}_{11}\chi_{\mathcal{F}}^{(12)}\\
\mathcal{A}_{11}\mathcal{B}_{12}\chi_{\mathcal{F}}^{(11)} & 1
\end{array}\right]\left(\begin{array}{c}
\varGamma_{1}b_{1,in}\\
\varGamma_{2}b_{2,in}
\end{array}\right)\nonumber \\
 &  & +\left[\begin{array}{c}
\mathcal{B}_{11}\chi_{\mathcal{F}}^{(12)}\\
\mathcal{B}_{12}\chi_{\mathcal{F}}^{(11)}
\end{array}\right]\mathcal{K}_{1}a_{1,in},
\end{eqnarray}
where $\chi_{B}=\text{diag}([\chi_{\mathcal{FF}}^{(11,12)}],[\chi_{\mathcal{FF}}^{(12,11)}])$
is the susceptibility matrix with $\chi_{\mathcal{FF}}^{(\alpha,\beta)}=\mathcal{F}\left[\mathcal{A}_{\alpha}\mathcal{B}_{\alpha}\mathcal{F}(\mathcal{A}_{\beta}\mathcal{B}_{\beta})\right]$
($\alpha,\beta=\{11,12\}$). In this case, the phonon number for MR1
is $\bar{n}_{1}\approx$ $\bar{m}_{1}\mathcal{T}_{11}+\bar{m}_{2}\mathcal{T}_{21}$,
with $\mathcal{T}_{1(2)1}$ $=$ $\frac{1}{2\pi}\int T_{R_{1(2)}\rightarrow b_{1}}(\omega)d\omega$,
$T_{R_{1}\rightarrow b_{1}}(\omega)=|\chi_{\mathcal{FF}}^{(11,12)}\mathit{\Gamma}_{1}|^{2}$,
and $T_{R_{2}\rightarrow b_{1}}(\omega)=|\mathcal{A}_{12}\mathcal{B}_{11}\chi_{\mathcal{F}}^{(12)}\chi_{\mathcal{FF}}^{(11,12)}\varGamma_{2}|^{2}$.
In comparison with the $a_{1}-b_{1}$ setup, $T_{R_{1}\rightarrow b_{1}}(\omega)$
is increased by a factor of $\chi_{\mathcal{F}}^{(12)}$ by coupling
$a_{1}$ to $b_{2}$. As shown in Figs. \ref{fig:model}(b) and \ref{fig:model}(c),
for $G_{12}=G_{11}$, the peak values of $T_{R_{1}\rightarrow b_{1}}$
(solid purple) and $T_{R_{2}\rightarrow b_{1}}$ (dashed green) are
equal to each other, and reach 1/4 of that for the MR1 solely in equilibrium
with $R_{1}$ (i.e. the red curve). As such, the net thermal noise
flow out of the MR1 reads $\delta n_{1}\approx-3\bar{m}_{1}/4+\bar{m}_{2}/4$,
which can not be effectively guided to the cavity mode and its zero-temperature
bath, so that the sideband cooling is inhibited, and vice versa for
the MR2.

\begin{figure*}[t]
\centering{}\includegraphics[width=1.5\columnwidth]{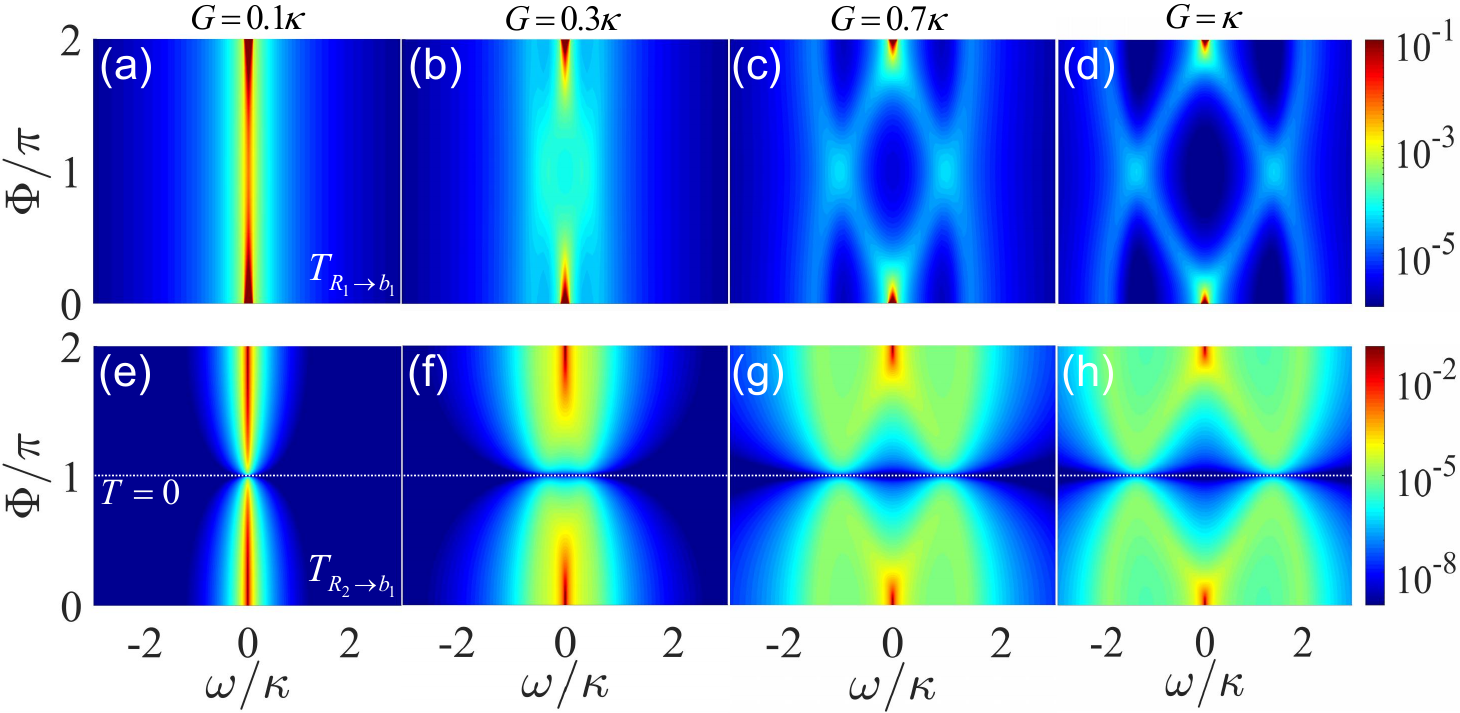}\caption{\label{fig:T33_T34}$T_{R_{1}\rightarrow b_{1}}(\omega)$ {[}(a)-(d){]}
and $T_{R_{2}\rightarrow b_{1}}(\omega)$ {[}(e)-(h){]} as functions
of the overall phase $\Phi/\pi$ and the frequency $\omega/\kappa$
for different coupling strengths $G/\kappa=0.1$ {[}(a), (e){]}, $G/\kappa=0.3$
{[}(b), (f){]}, $G/\kappa=0.7$ {[}(c), (g){]}, and $G/\kappa=1.0$
{[}(d), (h){]}. Here we set $G_{jk}=G$ ($j,k=1,2$). The white dotted
line indicates $T_{R_{2}\rightarrow b_{1}}$ $=$ $0$. Other parameters
are the same as in Fig. \ref{fig:model}(b).}
\end{figure*}

\textit{Gauge-invariant phase} - Now we discuss the four-mode configuration
in Fig. \ref{fig:model}(a). Under gauge transformation, the Hamiltonian
(\ref{eq:H_1}) can be rewritten as 
\begin{eqnarray}
H_{1} & = & \sum_{k=1,2}\Delta_{k}b_{k}^{\dagger}b_{k}+(G_{11}a_{1}b_{1}^{\dagger}+G_{12}a_{1}b_{2}^{\dagger}\nonumber \\
 &  & +G_{21}e^{-i\Phi}a_{2}b_{1}^{\dagger}+G_{22}a_{2}b_{2}^{\dagger}+\text{H.c.}),
\end{eqnarray}
where only the overall plaquette phase (or the loop phase) $\Phi=\phi_{11}+\phi_{21}-\phi_{12}-\phi_{22}$
is relevant to thermal flow in the plaquette. After eliminating the
optical degree of freedom, we can write the coupled equation for MRs
in the compact form (Appendix B)

\begin{gather}
\chi_{B}^{-1}(\Phi)\left(\begin{array}{c}
b_{1}\\
b_{2}
\end{array}\right)=\mathcal{H}(\Phi)\left(\begin{array}{c}
\varGamma_{1}b_{1,in}\\
\varGamma_{2}b_{2,in}
\end{array}\right)+\mathcal{M}(\Phi)\left(\begin{array}{c}
\mathcal{K}_{1}a_{1,in}\\
\mathcal{K}_{2}a_{2,in}
\end{array}\right),\label{eq:NoiseFlow_4ModeA}
\end{gather}
where the matrix elements of $\mathcal{H}(\Phi)$ are $\mathcal{H}_{11}=\mathcal{H}_{22}=1$,
$\mathcal{H}_{12}=(\mathcal{A}_{12}\mathcal{B}_{11}+e^{-i\Phi}\mathcal{A}_{22}\mathcal{B}_{21})\chi_{\mathcal{F}}^{(12,22)}$,
and $\mathcal{H}_{21}=(\mathcal{A}_{11}\mathcal{B}_{12}+e^{i\Phi}\mathcal{A}_{21}\mathcal{B}_{22})\chi_{\mathcal{F}}^{(11,21)}$.
The explicit forms of $\chi_{B}(\Phi)$ and $\mathcal{M}(\Phi)$ are
too cumbersome to be presented here. It shows that the thermal noise
bidirectionally flows between MR1 and MR2 through two paths $b_{2}\leftrightarrow a_{1}(a_{2})\leftrightarrow b_{1}$,
which are superimposed with each other and are dependent on the overall
phase $\Phi$. In this case, the net thermal noise flow is given by
$\delta n_{1(2)}=$ $\bar{m}_{1(2)}\left(\mathcal{T}_{11(22)}-1\right)+\bar{m}_{2(1)}\mathcal{T}_{21(12)}$.
To achieve ground-state cooling of the two MRs, we must maximize the
noise flow out of the MR itself {[}i.e. $\bar{m}_{1(2)}(\mathcal{T}_{11(22)}-1$){]},
and meanwhile significantly reduce or, ideally, completely eliminate
the flow of thermal noise from the other {[}i.e. $\bar{m}_{2(1)}\mathcal{T}_{21(12)}${]},
particularly the one that is exposed to a thermal bath with a larger
thermal phonon number. We then consider the destructive interference
phase, for which $\mathcal{H}_{12}=\mathcal{H}_{21}=0$ and therefore
the bidirectional noise flow is suppressed. It follows that
\begin{equation}
\text{Exp}(i\Phi)=-\frac{G_{21}G_{22}\chi_{a2}}{G_{11}G_{12}\chi_{a1}}=-\frac{G_{11}G_{12}\chi_{a1}}{G_{21}G_{22}\chi_{a2}},\label{eq:Matching_condition}
\end{equation}
 which imposes the conditions $\Phi=\pi$ and
\begin{equation}
\frac{G_{11}G_{12}}{\kappa_{1}/2-i\omega}=\frac{G_{21}G_{22}}{\kappa_{2}/2-i\omega}.\label{eq:Interference_condition}
\end{equation}
Remarkably, this condition involves both the coupling strengths $G_{jk}$
and the cavity dissipation $\kappa_{j}$ (but not mechanical damping)
for optomechanical interfaces, and thus, the underlying physics can
not be simply described by coherent control theory. For $\Phi=\pi$,
Hamiltonian (\ref{eq:H_1}) can be re-organized as
\[
H_{1}=\sum_{k=1,2}\Delta_{k}b_{k}^{\dagger}b_{k}+(G_{1}b_{1}\alpha_{1,-}^{\dagger}+G_{2}b_{2}\alpha_{2,+}^{\dagger}+\text{H.c.}),
\]
 with $G_{k}=\sqrt{G_{1k}^{2}+G_{2k}^{2}}$ and
\begin{align}
\alpha_{1,+}\propto G_{21}a_{1}+G_{11}a_{2}, & \text{ }\alpha_{1,-}\propto G_{11}a_{1}-G_{21}a_{2},\nonumber \\
\alpha_{2,+}\propto G_{12}a_{1}+G_{22}a_{2}, & \text{ }\alpha_{2,-}\propto G_{22}a_{1}-G_{12}a_{2}.
\end{align}
Moreover, for $\kappa_{1}=\kappa_{2}$ the destructive interference
condition reduces to $G_{11}G_{12}=G_{21}G_{22}$, giving rise to
$\alpha_{2,\pm}=(G_{12}/G_{21})\alpha_{1,\pm}$, where both the cavity
supermodes $\alpha_{1,+}$ and $\alpha_{2,+}$ are orthogonal to $\alpha_{1(2),-}$.
As such, the two MRs can be simultaneously cooled through beam splitter
interactions $G_{1}b_{1}\alpha_{-}^{\dagger}$ and $G_{2}b_{2}\alpha_{+}^{\dagger}$
(where $\alpha_{\pm}\sim\alpha_{1(2),\pm}$), without exchanging thermal
phonons.

\textit{Results} - To gain an intuitive understanding, we first assume
that the coupling strengths are identical, i.e. $G_{jk}=G$ ($j,k=1,2$),
so that $\alpha_{1\pm}=\alpha_{2\pm}=(a_{1}\pm a_{2})/\sqrt{2}$.
Fig. \ref{fig:T33_T34} shows the scattering coefficients $T_{R_{1}\rightarrow b_{1}}(\omega)$
{[}Figs. \ref{fig:T33_T34}(a)-(d){]} and $T_{R_{2}\rightarrow b_{1}}(\omega)$
{[}Figs. \ref{fig:T33_T34}(e)-(h){]} from the own heat bath $R_{1}$
and from the bath $R_{2}$, respectively, which is exactly the same
to $T_{R_{2}\rightarrow b_{2}}(\omega)$ and $T_{R_{1}\rightarrow b_{2}}(\omega)$
for MR2. Under the weak coupling condition $G/\kappa=0.1$ and setting
$\Delta_{k}=0$ for simplicity (we have discussed the influence of
laser detunings in Appendix C), both $T_{R_{1}\rightarrow b_{1}}(\omega)$
and $T_{R_{2}\rightarrow b_{1}}(\omega)$ {[}as shown by Fig. \ref{fig:T33_T34}(a)
and Fig. \ref{fig:T33_T34}(e){]} have a unique peak centered at $\omega=0$
and gradually decrease when $\Phi$ is steered towards $\pi$, and
remarkably, $T_{R_{2}\rightarrow b_{1}}(\omega=0)$ vanishes at $\Phi=\pi$.
As the coupling strengths increase, we find that $T_{R_{1}\rightarrow b_{1}}(\omega)$
is split into four peaks for $\Phi\neq0$, which merge into two peaks
specially for $\Phi=\pi$ due to the couplings with the two degenerated
superposition (normal) modes $\alpha_{1\pm}(\alpha_{2\pm})$, see
Figs. \ref{fig:T33_T34}(c) and \ref{fig:T33_T34}(d) with $G/\kappa=0.7$
and $G/\kappa=1$. It is interesting to see that the four peaks in
$T_{R_{2}\rightarrow b_{1}}(\omega)$ do not merge at $\Phi=\pi$,
but instead, there appears two avoided crossings around $\Phi=\pi$
due to destructive interference between the multiple coupling channels,
as shown by Fig. \ref{fig:T33_T34}(g) {[}\ref{fig:T33_T34}(h){]}.
As a result, the thermal noise flow between the two MRs can be completely
eliminated (i.e. $\bar{m}_{2}\mathcal{T}_{21}=0$ and $\bar{m}_{1}\mathcal{T}_{12}=0$)
in both the weak and strong coupling regimes.

\begin{figure}[t]
\centering{}\includegraphics[width=1\columnwidth]{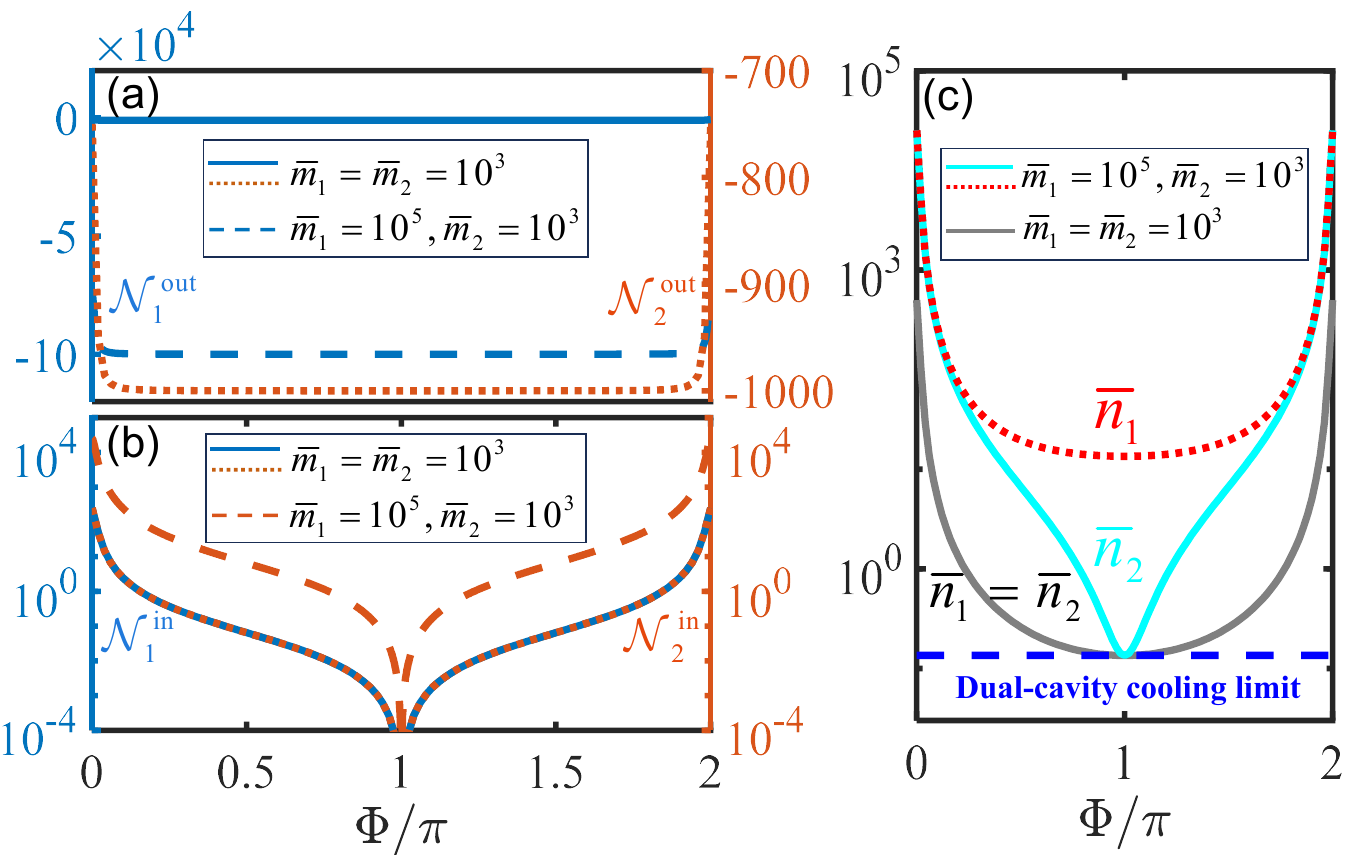}\caption{\label{fig:nf}(a) $\mathcal{N}_{1(2)}^{\text{out}}$, (b) $\mathcal{N}_{1(2)}^{\text{in}}$,
and (c) phonon occupation numbers $\bar{n}_{1(2)}$ as functions of
the overall phase $\Phi/\pi$, where the thermal phonon number for
the baths are chosen as $\bar{m}_{1}=\bar{m}_{2}$ $=$ $10^{3}$
and $\bar{m}_{1}=10^{5}$, $\bar{m}_{2}=$ $10^{3}$, respectively.
The horizontal dashed line in (c) indicates the dual-cavity cooling
limit. Other parameters are the same as in Fig. \ref{fig:model}(b).}
\end{figure}

In Fig. \ref{fig:nf}, by first considering the weak coupling regime
with $G/\kappa=0.1$ and the thermal phonon number $\bar{m}_{1}=\bar{m}_{2}=10^{3}$
(or $\bar{m}_{1}=10^{5},\bar{m}_{2}=10^{3}$), we study the flow of
thermal noises $\delta n_{1(2)}$ $=$ $\bar{m}_{1(2)}(\mathcal{T}_{11(22)}-1)$
$+$ $\bar{m}_{2(1)}\mathcal{T}_{21(12)}$ and phonon occupation numbers
$\bar{n}_{1(2)}$ for the two MRs. We show the thermal noise flow
out of MR1 (MR2) $\mathcal{N}_{1(2)}^{\text{out}}=\bar{m}_{1(2)}(\mathcal{T}_{11(22)}-1)$
in Fig. \ref{fig:nf}(a) and the noise inflow from the other {[}i.e.
MR2 (MR1){]} $\mathcal{N}_{1(2)}^{\text{in}}=\bar{m}_{2(1)}\mathcal{T}_{21(12)}$
in Fig. \ref{fig:nf}(b). We find $\mathcal{N}_{1(2)}^{\text{out}}$
$\approx$$-3\bar{m}_{1(2)}/4$ and $\mathcal{N}_{1(2)}^{\text{in}}$
$\approx$ $\bar{m}_{1(2)}/4$ for $\Phi=0$, which are the same to
those in the case of the $b_{1}-a_{1}-b_{2}$ setup. For $\Phi\neq0$,
the simultaneous ground-state cooling of the two MRs (i.e. $\bar{n}_{1}<1$
and $\bar{n}_{2}<1$) can be achieved for $\Phi\in(0.23\pi,1.77\pi)$,
see the gray solid curve in Fig. \ref{fig:nf}(c). In particular,
for $\Phi=\pi$, $\mathcal{N}_{1(2)}^{\text{out}}$ and $\mathcal{N}_{1(2)}^{\text{in}}$
reach the minimum and are given by $\mathcal{N}_{1(2)}^{\text{out}}$
$\approx$ $-\bar{m}_{1(2)}$ and $\mathcal{N}_{1(2)}^{\text{in}}$
$=$ $0$. When the phonon number of the thermal baths $R_{1}$ and
$R_{2}$ are both equal to $\bar{m}_{1}=\bar{m}_{2}=10^{3}$, the
thermal noise flow between the two MRs is reciprocal, i.e. $\mathcal{N}_{2}^{\text{out}}=\mathcal{N}_{1}^{\text{out}}$
and $\mathcal{N}_{2}^{\text{in}}=\mathcal{N}_{1}^{\text{in}}$ as
denoted by the solid and dotted curves in Figs. \ref{fig:nf}(a) and
\ref{fig:nf}(b). We find that the phonon occupation numbers $\bar{n}_{1}=\bar{n}_{2}\approx0.135$
are achieved, see the gray curve in Fig. \ref{fig:nf}(c). If one
of the MRs is suffering from a ``hotter'' reservoir, for example,
when the thermal phonon number of $R_{1}$ is increased to $\bar{m}_{1}=10^{5}$
{[}see the blue dashed curve in Fig. \ref{fig:nf}(a){]}, $\mathcal{N}_{1}^{\text{out}}\approx-10^{5}$
has a relatively large minimum and $\mathcal{N}_{1}^{\text{in}}=0$
for $\text{\ensuremath{\Phi=\pi}}$, but the ground-state cooling
of MR1 cannot be achieved {[}see the red dotted curve in Fig. \ref{fig:nf}(c){]}.
While for MR2, the noise flow $\mathcal{N}_{2}^{\text{in}}$ from
$R_{1}$ increases correspondingly for $\text{\ensuremath{\Phi\neq\pi}}$,
even a single thermal phonon flow into MR2 {[}i.e. $\mathcal{N}_{2}^{\text{in}}>1$,
see the red dashed curve Fig. \ref{fig:nf}(b){]} can hinder ground-state
cooling of MR2. Nevertheless, for $\text{\ensuremath{\Phi}}$ around
$\pi$, the phonon number of MR2 $\bar{n}_{2}<1$ can still be achieved,
and its minimum reaches the dual-cavity cooling limit (indicated by
the blue dashed line) due to the complete elimination of $\mathcal{N}_{2}^{\text{in}}$.
\begin{figure}[H]
\centering{}\includegraphics[width=1\columnwidth]{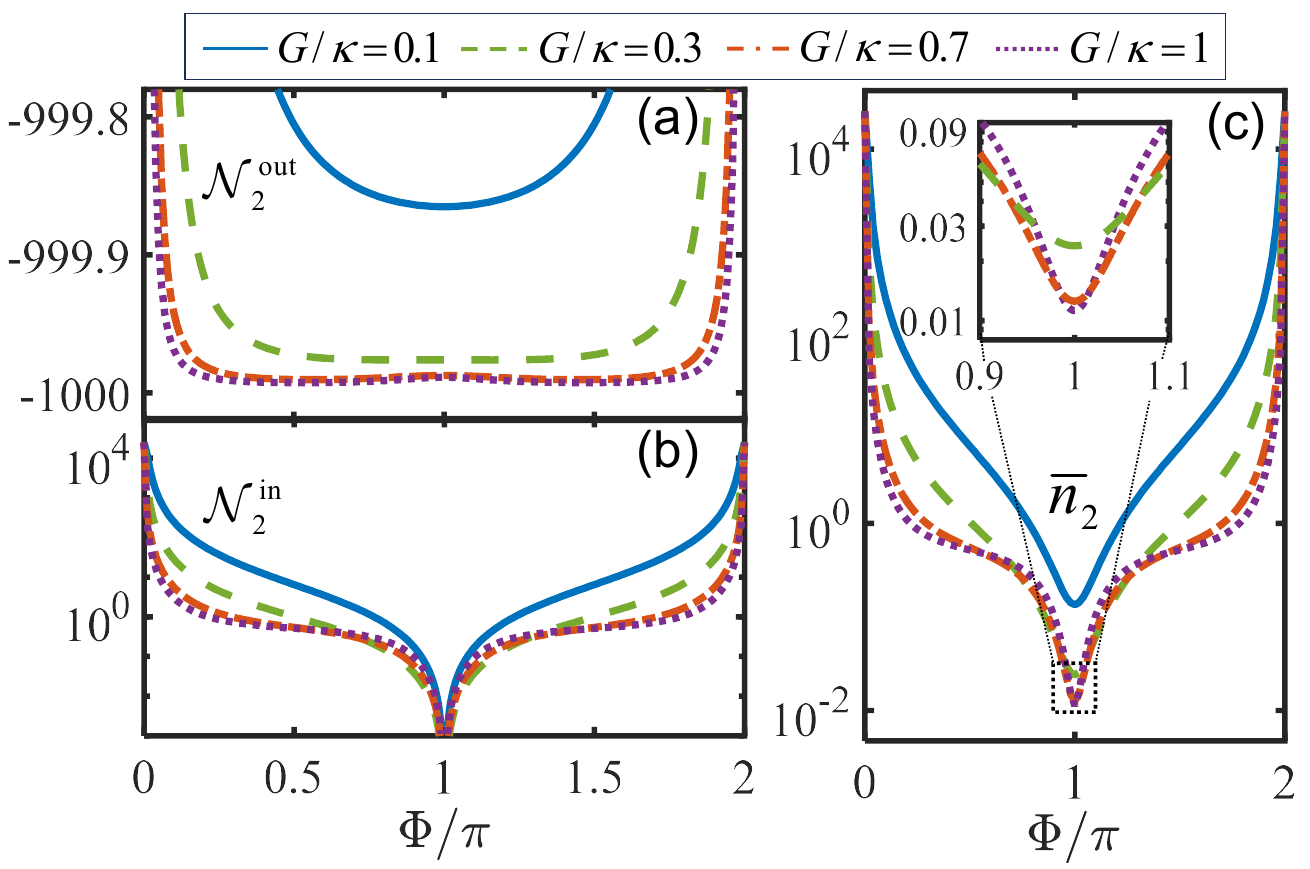}\caption{\label{fig:4_vs_G}(a) $\mathcal{N}_{2}^{\text{out}}$, (b) $\mathcal{N}_{2}^{\text{in}}$,
and (c) phonon occupation numbers $\bar{n}_{2}$ as functions of the
overall relative phase $\Phi/\pi$ in different coupling strength
$G/\kappa=0.1$ (solid curve), $G/\kappa=0.3$ (dashed curve), $G/\kappa=0.7$
(dashed-dotted curve), and $G/\kappa=1.0$ (dotted curve), where the
thermal noises of $R_{1}$ and $R_{2}$ are $\bar{m}_{2}=10^{3}$
and $\bar{m}_{1}=10^{5}$. Other parameters are the same as in Fig.
\ref{fig:model}(b).}
\end{figure}

We then consider the intermediate and strong coupling regimes by considering
the set of coupling strengths $G/\kappa=\{0.1,0.3,0.7,1\}$. As $G$
becomes comparable with the cavity linewidth $\kappa$ (i.e. the strong
coupling regime $G/\kappa\sim1$), the $\Phi$-dependent noise flow
out of MR2 $\mathcal{N}_{2}^{\text{out}}$ displays a double-well
lineshape, where the local minima are found at around $\Phi=\pi/2$
and $\Phi=3\pi/2$, see Fig. \ref{fig:4_vs_G}(a). However, since
the thermal phonon exchange around $\Phi=\pi/2,3\pi/2$ can not be
fully suppressed, the ground-state cooling of MR2 could not be possible
if far more than one thermal phonon from $R_{1}$ flows into MR2 {[}i.e.
$\mathcal{N}_{2}^{\text{in}}\gg1$, see Fig. \ref{fig:4_vs_G}(b){]}.
In comparison, although $\mathcal{N}_{2}^{\text{out}}$ does not reach
minimal at $\Phi=\pi$, the best cooling efficiency of MR2 is still
attainable here due to $\mathcal{N}_{1(2)}^{\text{in}}=0$. Fig. \ref{fig:4_vs_G}(c)
shows $\bar{n}_{2}$ versus $\Phi$ with $\bar{m}_{2}=10^{3}$ and
$\bar{m}_{1}=10^{5}$, where the minimum of $\bar{n}_{2}$ saturates
to 0.01 as $G/\kappa\rightarrow1$.

\begin{figure}[t]
\centering{}\includegraphics[width=1\columnwidth]{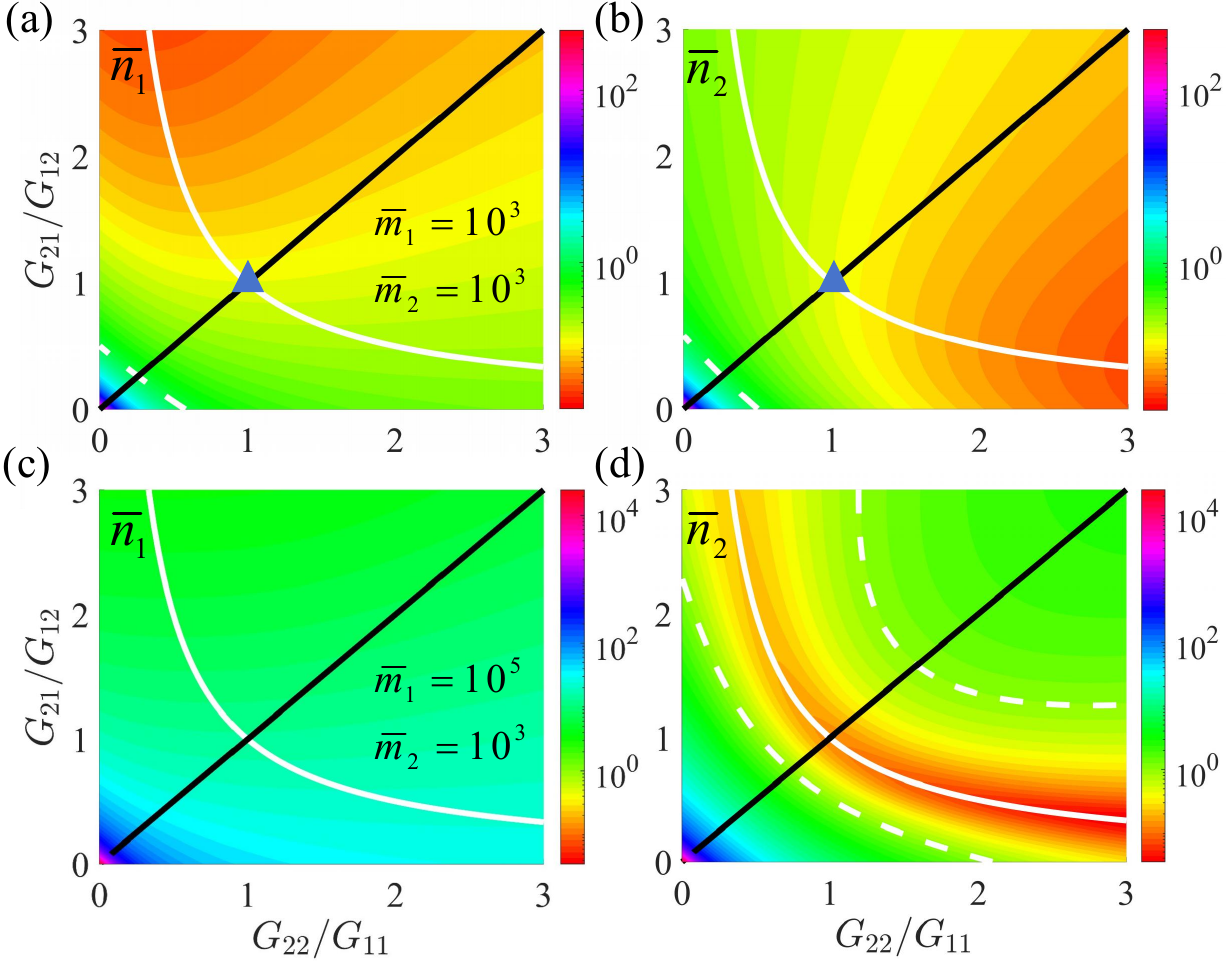}\caption{\label{fig:n_G22_G21}Phonon occupation numbers $\bar{n}_{1}$ {[}(a)
and (c){]} and $\bar{n}_{2}$ {[}(b) and (d){]} as functions of the
coupling strength $G_{22}/G_{11}$ and $G_{21}/G_{12}$ in different
phonon number of the thermal noise {[}(a), (b){]}, $\bar{m}_{1}=10^{3}$,
$\bar{m}_{2}=10^{3}$ and {[}(c), (d){]}, $\bar{m}_{1}=10^{5}$, $\bar{m}_{2}=10^{3}$.
The white dashed contour correspond to $\bar{n}_{k}=1$ ($k$ $\in$
$\{1,2\}$). The white and black solid curves denote the conditions
$(G_{22}/G_{11})^{-1}$ $=G_{21}/G_{12}$ and $G_{22}/G_{11}$ $=G_{21}/G_{12}$,
respectively, and their intersection points are marked by the dark
blue triangle. In all figures we assume $G_{11}/\kappa$ $=G_{12}/\kappa$
$=0.1$ and $\Phi=\pi$. Other parameters are the same as in Fig.
\ref{fig:model}(b).}
\end{figure}

When the coupling strengths are not fully the same, the destructive
interference condition {[}Eq. (\ref{eq:Interference_condition}){]}
becomes $(G_{22}/G_{11})^{-1}=G_{21}/G_{12}$ for $\kappa_{1}=\kappa_{2}$,
which corresponds to the impedance matching condition with a flat
band (i.e., independent of $\omega$). We focus on the phonon occupation
number $\bar{n}_{1(2)}\approx$ $\bar{m}_{1(2)}\mathcal{T}_{11(22)}+\bar{m}_{2(1)}\mathcal{T}_{21(12)}$
versus the coupling strength $G_{22}/G_{11}$ and $G_{21}/G_{12}$
in Fig. \ref{fig:n_G22_G21}, by considering the thermal baths with
$\bar{m}_{1}=\bar{m}_{2}=10^{3}$ {[}Figs. \ref{fig:n_G22_G21}(a)
and \ref{fig:n_G22_G21}(b){]}. The destructive interference regime
is indicated by the white solid lines. For comparison, we also indicate
the regime $G_{22}/G_{11}=G_{21}/G_{12}$ with the black solid curves,
which is referred to as the dark-mode breaking condition since both
the mechanical superimpose modes (the dark and bright modes) are individually
coupled to the cold reservoir ($a_{1}$ and $a_{2}$) \citep{Liu2022a,Huang2022}.
As can be seen in Figs. \ref{fig:n_G22_G21}(a) and \ref{fig:n_G22_G21}(b),
for a given $G_{21}/G_{12}$ ($G_{22}/G_{11}$) and as $G_{22}/G_{11}$
($G_{21}/G_{12}$) increases, $\bar{n}_{1}$ ($\bar{n}_{2}$) reaches
its minimum {[}denoted as $\bar{n}_{1}^{(min)}$ ($\bar{n}_{2}^{(min)}$){]}
whenever $(G_{22}/G_{11})^{-1}=G_{21}/G_{12}$. Specifically, one
has $\bar{n}_{1}^{(min)}=\bar{n}_{2}^{(min)}$ for $G_{21}/G_{12}=G_{22}/G_{11}=1$
(marked by the triangles), where the dual-cavity cooling limits for
the two MRs are the same. The minimum phonon number in this regime
{[}$G_{21}/G_{12}=(G_{22}/G_{11})^{-1}${]} are $\bar{n}_{1}^{(min)}=\bar{m}_{1}\mathcal{T}_{11}$
and $\bar{n}_{2}^{(min)}=\bar{m}_{2}\mathcal{T}_{22}$ since the noise
inflow from the other $\mathcal{N}_{1(2)}^{\text{in}}$ vanish. Moreover,
$\bar{n}_{1}^{(min)}$ ($\bar{n}_{2}^{(min)}$) decreases with the
increase of $G_{21}/G_{12}$ ($G_{22}/G_{11}$) {[}i.e. the optomechanical
coupling strengths between MR1 (MR2) and the cavity mode $a_{2}${]}
for a given $G_{22}/G_{11}$ ($G_{21}/G_{12}$).

However, if the MRs' thermal baths have different thermal phonon number
$\bar{m}_{1}=10^{5}$ and $\bar{m}_{2}=10^{3}$ {[}Figs. \ref{fig:n_G22_G21}(c)
and \ref{fig:n_G22_G21}(d){]}, the thermal noise flowing into the
MR1 will grow increasing to a higher temperature. As a result, MR1
fails to achieve ground-state cooling for $\bar{m}_{1}=10^{5}$, and
meanwhile, the MR2 can be heated up if the condition $(G_{22}/G_{11})^{-1}=G_{21}/G_{12}$
is not satisfied, then the parameter region (bounded by the white
dashed lines) for realizing ground-state cooling of MR2 substantially
shrinks. Nevertheless, since $\mathcal{N}_{2}^{\text{in}}=0$ for
$(G_{22}/G_{11})^{-1}=G_{21}/G_{12}$, the phonon occupation number
of MR2 $\bar{n}_{2}\approx\bar{m}_{2}\mathcal{T}_{22}$ is independent
of $\bar{m}_{1}$, preserving the cooling performance for MR2 from
the larger thermal noise {[}see the white solid line in Fig. \ref{fig:n_G22_G21}(d){]}.
In this circumstance, $\bar{n}_{2}$ may no longer monotonically decrease
with the increase of $G_{22}/G_{11}$ for a sufficiently large $G_{21}/G_{12}$.
Although MR1 can not be cooled to the ground state here, its effective
temperature still achieves dual-cavity cooling limit when the $(G_{22}/G_{11})^{-1}=G_{21}/G_{12}$
is met.

\begin{figure}[t]
\centering{}\includegraphics[width=1\columnwidth]{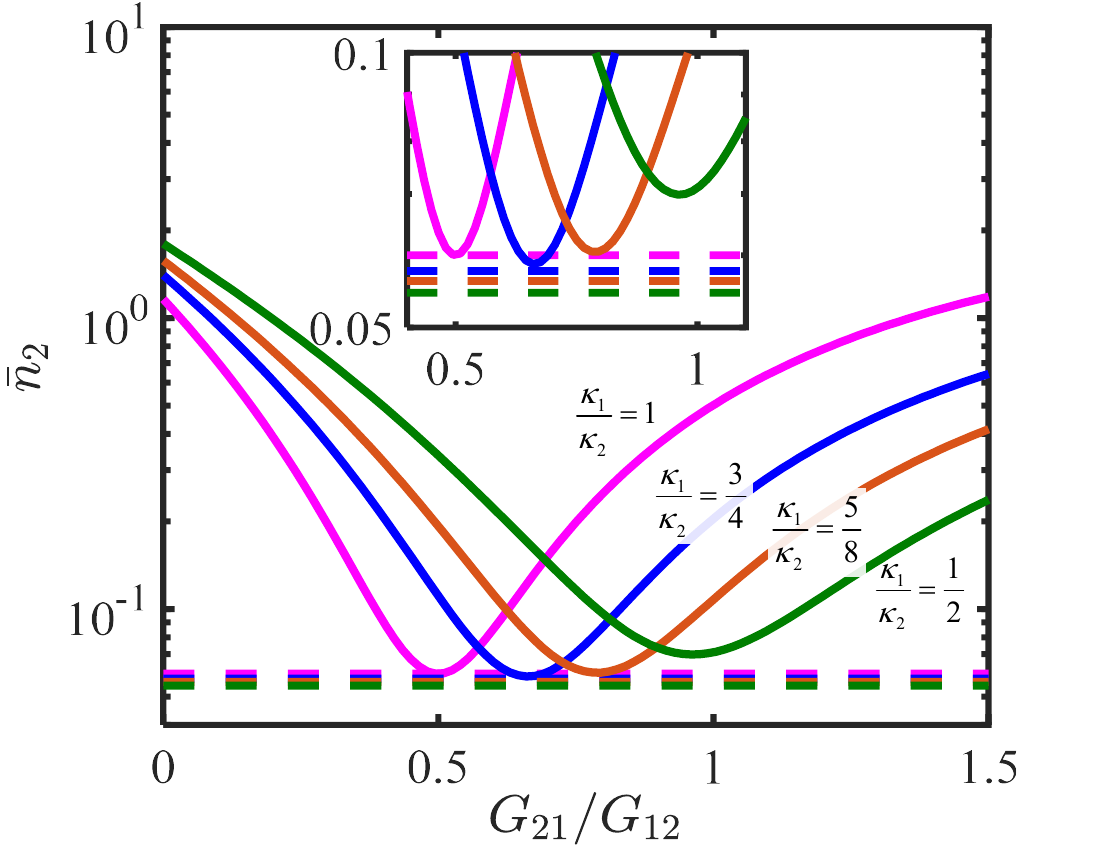}\caption{\label{fig:n_G21 in diff kappa}Phonon occupation number $\bar{n}_{2}$
as a function of the coupling strength $G_{21}/G_{12}$ with $G_{22}/G_{11}=2$
in a set of different ratios of the decay rates: $\kappa_{1}/\kappa_{2}=$
$\{1,3/4,5/8,1/2\}$ with $\kappa_{2}=\kappa$. The horizontal dashed
curves correspond to the dual-cavity cooling limits for MR2. Other
parameters are the same as in Fig. \ref{fig:n_G22_G21}.}
\end{figure}

Previously, we concentrated on scenarios where $\kappa_{1}=\kappa_{2}$.
Moving forward, we consider the case of $\kappa_{1}\neq\kappa_{2}$.
For $\gamma_{1(2)}\ll\kappa_{1},\kappa_{2}$, it is effective to consider
a narrow bandwidth of the response function $\chi_{aj}$ around $\omega=0$
which fulfills $|\delta\omega|\ll\kappa_{1(2)}$. Then, Eq. (\ref{eq:Interference_condition})
can be approximately rewritten by
\begin{equation}
\frac{G_{11}}{G_{22}}\approx\frac{G_{21}}{G_{12}}\frac{\kappa_{1}}{\kappa_{2}}.\label{eq:IM_Kappa}
\end{equation}
It reveals that the impedance matching condition for the coupling
strengths now depends on the ratio $\kappa_{1}/\kappa_{2}$, which
typically varies for different experimental sampling devices. In Fig.
\ref{fig:n_G21 in diff kappa}, we show the phonon occupation number
$\bar{n}_{2}$ as a function of $G_{21}/G_{12}$ by setting $\kappa_{1}/\kappa_{2}$
$=$ $\{1,3/4,5/8,1/2\}$ and $G_{22}/G_{11}=2$. Obviously, the valleys
indicating the optimal cooling performance appear around the coupling
strengths $G_{21}/G_{12}$ $=$ $\{1/2,2/3,4/5,1\}$ corresponding
to the fulfillment of Eq. (\ref{eq:IM_Kappa}). Moreover, the minimum
phonon number are very close to those at the dual-cavity cooling limit,
see the dashed lines. As $\kappa_{1}$ and $\kappa_{2}$ differ from
each other by 100\% (the green curve), the influence of the finite
bandwidth $\delta\omega$ on the transmission coefficient $T_{R_{1}\rightarrow b_{2}}(\omega)$
can be prominent, see the gap between the valley and the dual-cavity
cooling limit, as also shown in the inset of Fig. \ref{fig:n_G21 in diff kappa}.

\textit{Further discussion and conclusion }- The proposed scheme can
be implemented in a four-mode optomechanical setup, which has been
experimentally demonstrated in the microwave circuit \citep{Xu2016,Damskagg2019}.
Considering the set of parameters related to Refs. \citep{Xu2016,Damskagg2019,Peterson2017,Yang2020,Cao2025}:
the damping rates $(\gamma,\kappa)/2\pi$ $\sim$ $(1\text{ Hz},200\text{ kHz})$,
the mechanical frequencies $\omega_{b,1}/2\pi=1$ MHz and $\omega_{b,2}/2\pi=5$
MHz, then the phonon number $\bar{m}_{1}$ $\sim$ $10^{3}$ (or $\bar{m}_{1}\sim10^{5}$)
and $\bar{m}_{2}$ $\sim$ $10^{3}$ correspond to the thermal temperature
$T_{1}=40$ mK (or $T_{1}=4$ K) and $T_{2}=240$ mK. Alternatively,
for the parameters $(\gamma,\kappa)/2\pi$ $\sim$ $(10\text{ Hz},1\text{ MHz})$,
($\omega_{b,1},\omega_{b,2}/)/2\pi=(5,10)$ MHz, the phonon number
$\bar{m}_{1}$ $\sim$ $10^{5}$ and $\bar{m}_{2}$ $\sim$ $10^{3}$
corresponds to the thermal temperature $T_{1}=20$ K and $T_{2}=500$
mK. The parameters above both ensure that the system is well in the
resolved sideband regime. Thus, by engineering the thermal noise flow,
the MRs {[}with $(\omega_{b,1},\omega_{b,2})/2\pi=(6.7,9.4)$ MHz{]}
can be cooled down to the ground state ($\bar{n}_{1}\approx$0.216,
$\bar{n}_{2}\approx$0.1485) for $T_{1}=T_{2}\sim500$ mK, and $\bar{n}_{1}\approx$8.37,
$\bar{n}_{2}\approx$0.1485 even though the MRs are in largely different
cryogenic environment ($T_{1}=20$ K and $T_{2}=500$ mK).

In conclusion, we have studied the thermal noise flow in a four-mode
optomechanical plaquette, where the two non-degenerate MRs are subject
to the thermal environments with different thermal phonon numbers
or thermal temperatures. By engineering the optomechanical coupling
strengths and phases, we show that the thermal noise flow between
the MRs can be fully suppressed for the overall phase $\Phi\sim\pi$,
despite a temperature difference for the individual thermal environment.
As a result, for a lower cryogenic temperature of a few hundred mK,
the simultaneous ground-state cooling of the two MRs towards the dual-cavity
cooling limit can be realized; while for one of the MRs subjected
to the ``hot'' reservoir with a high temperature (e.g. a few tens
of Kelvin), the MR subjected to the low-temperature bath can still
be effectively cooled without being heated up by the other bath. The
proposed method depends on both the coherent optomechanical coupling
and cavity dissipation, and is robust to variations in laser detunings
(Appendix C) and cavity decay rates. Moreover, it can be applied to
manipulate thermal noise flow in a nonreciprocal way, see Appendix
D. 
It expands the scope of control from purely coherent dynamics to include the manipulation of thermal noise, thereby opening new avenues for managing energy flow in multimode quantum systems and building optomechanical networks in complicated thermal environments.

Acknowledgements - The authors thank Jiteng Sheng for helpful comments
and suggestions. H.W. acknowledges support from the National Natural
Science Foundation of China under Grant No. 12174058. Y.L. was supported
by the National Natural Science Foundation of China under Grant No.
12274107 and the Research Funds of Hainan University {[}Grant No.
KYQD(ZR)23010{]}.





\bibliographystyle{apsrev4-1}
\bibliography{My-Collection}

\newpage{}

\begin{widetext}

\section*{Appendix A Linearized Hamiltonian}

The Hamiltonian of the four-mode optomechanical system is ($\hbar=1$)
\begin{equation}
H=\sum_{j=1,2}\omega_{a,j}a_{j}^{\dagger}a_{j}+\sum_{k=1,2}\omega_{b,k}b_{k}^{\dagger}b_{k}+\sum_{j,k}g_{j,k}a_{j}^{\dagger}a_{j}(b_{k}^{\dagger}+b_{k}),
\end{equation}
where $(a_{j}$, $a_{j}^{\dagger})_{j=1,2}$ are the annihilation
and creation operators of the cavity modes; $(b_{k}$, $b_{k}^{\dagger})_{k=1,2}$
are the operators of the mechanical modes. The parameter $g_{j,k}$
is the single photon coupling strength between the cavity mode $a_{j}$
and the mechanical mode $b_{k}$. Such a model has been realized in
microwave- or electro-optomechanical systems \citep{Xu2016,Weaver2017,Peterson2017,Damskagg2019,Yang2020,Lake2020,Cao2025}.
Furthermore, we consider the driving scheme, where the cavity mode
$a_{j}$ is driven by a two-tone laser at frequencies $\omega_{a,j}-\omega_{b,k}+\Delta_{j,k}$
\citep{Xu2016,Damskagg2019},
with the detunings $\Delta_{j,k}$ being tuned around the mechanical
red sidebands, i.e. $\Delta_{j,k}$ $\ll$ $\{\omega_{b,k}$, $|\omega_{b,1}-\omega_{b,2}|\}$.
We assume that min$[\omega_{b,k}$, $|\omega_{b,1}-\omega_{b,2}|]$
$\gg$ max$[|g_{j,k}\alpha_{j,k}(t)|]$, then the coherent part can
be approximately given by $\alpha_{j}(t)$ $\approx$ $\sum_{k=1,2}\alpha_{j,k}e^{-i(\omega_{b,k}-\Delta_{j,k})t}$,
where $\alpha_{j,k}$ $=$ $|\alpha_{j,k}|e^{i\phi_{j,k}}$ are complex
numbers with $\phi_{j,k}$ relying on the phase of the corresponding
laser pump. For convenience, we first assume that the detunings $\Delta_{1,1}=\Delta_{2,1}=\Delta_{1}$
and $\Delta_{1,2}=\Delta_{2,2}=\Delta_{2}$. Then we can obtain the
linearized Hamiltonian, which in the frame rotating with $\sum_{j=1,2}\omega_{a,j}a_{j}^{\dagger}a_{j}$
$+$ $\sum_{k=1,2}(\omega_{b,k}-\Delta_{k})b_{k}^{\dagger}b_{k}$
reads
\begin{eqnarray}
H_{lin,1} & = & \Delta_{1}b_{1}^{\dagger}b_{1}+\Delta_{2}b_{2}^{\dagger}b_{2}+(G_{11}a_{1}b_{1}^{\dagger}+G_{12}a_{1}b_{2}^{\dagger}\nonumber \\
 &  & +G_{21}e^{-i\Phi}a_{2}b_{1}^{\dagger}+G_{22}a_{2}b_{2}^{\dagger}+\text{H.c.}),\label{eq:H_lin1}
\end{eqnarray}
where $G_{j,k}=|g_{j,k}\alpha_{j,k}|$ are the effective optomechanical
coupling strengths and the terms oscillating at frequencies close
to $|\omega_{b,1}\pm\omega_{b,2}|$ and $2\omega_{b,k}$ are neglected
under the rotating wave approximation (RWA). Moreover, we have made
a gauge transformation to the operators $a_{1}$ $\rightarrow$ $a_{1}e^{i\phi_{12}}$,
$a_{2}$ $\rightarrow$ $a_{2}e^{i\phi_{22}}$ and $b_{1}$ $\rightarrow$
$b_{1}e^{-i(\phi_{11}-\phi_{12})}$, and introduced a overall plaquette
phase $\Phi$ $=$ $\phi_{21}-\phi_{22}-(\phi_{11}-\phi_{12})$ to
the gauge system, which can be addressed using phase-correlated lasers.

\section*{Appendix B Input-output relations in frequency spectrum}

We can write the linearized quantum Langevin equations (QLEs) from
the Eq. (\ref{eq:H_lin1}) as
\begin{eqnarray}
\dot{a}_{1} & = & -i\left(G_{11}b_{1}+G_{12}b_{2}\right)-\frac{\kappa_{1}}{2}a_{1}+\sqrt{\kappa_{1}}a_{1,in},\nonumber \\
\dot{a}_{2} & = & -i\left(G_{21}b_{1}e^{i\Phi}+G_{22}b_{2}\right)-\frac{\kappa_{2}}{2}a_{2}+\sqrt{\kappa_{2}}a_{2,in},\nonumber \\
\dot{b}_{1} & = & -i\left(G_{11}a_{1}+G_{21}a_{2}e^{-i\Phi}\right)-(i\Delta_{b1}+\frac{\gamma_{1}}{2})b_{1}+\sqrt{\gamma_{1}}b_{1,in},\nonumber \\
\dot{b}_{2} & = & -i\left(G_{12}a_{1}+G_{22}a_{2}\right)-(i\Delta_{b2}+\frac{\gamma_{2}}{2})b_{2}+\sqrt{\gamma_{2}}b_{2,in},\label{eq:QLEs}
\end{eqnarray}
or in a compact form $\dot{v}(t)=Mv(t)+\sqrt{\Gamma}v_{in}(t)$, where
$v(t)$ $=$ $(a_{1}$, $a_{2}$, $b_{1}$, $b_{2})^{T}$ is the fluctuation
operator vector, $v_{in}(t)$ $=$ $(a_{1,in}$, $a_{2,in}$, $b_{1,in}$,
$b_{2,in})^{T}$ is the noise operator vector, $\Gamma$ $=$ diag$(\kappa_{1}$,
$\kappa_{2}$, $\gamma_{1}$, $\gamma_{2})$ is the diagonal damping
matrix, and the dynamical matrix 
\begin{eqnarray}
M & = & \left[\begin{array}{cccc}
-\frac{\kappa_{1}}{2} & 0 & -iG_{11} & -iG_{12}\\
0 & -\frac{\kappa_{2}}{2} & -iG_{21}e^{i\Phi} & -iG_{22}\\
-iG_{11} & -iG_{21}e^{-i\Phi} & -\frac{\gamma_{1}}{2}-i\Delta_{1} & 0\\
-iG_{12} & -iG_{22} & 0 & -\frac{\gamma_{2}}{2}-i\Delta_{2}
\end{array}\right].
\end{eqnarray}
The operators $a_{j,in}$ and $b_{k,in}$ are the noise operators
denoting the Gaussian white noises for the cavity modes and the mechanical
modes.

We then convert the QLEs into the frequency domain by applying the
Fourier transformation $o(\omega)=\frac{1}{\sqrt{2\pi}}\int o(t)e^{i\omega t}dt$
to arbitrary optical and mechanical (noise) operators $o(t)$, leading
to
\begin{equation}
v(\omega)=U(\omega)v_{in}(\omega),\label{eq:flow}
\end{equation}
where transformation matrix $U(\omega)=(-M-i\omega I)^{-1}\sqrt{\Gamma}$
and $I$ is the identity matrix. Eq. (\ref{eq:flow}) indicates that
input vacuum and thermal noises entering each mode through the corresponding
matrix element $U_{lm}(\omega)$, with $\{l,m\}$ $\in$ $\{1,2,3,4\}$
and can be re-organized as
\begin{eqnarray}
v(\omega) & = & U_{\chi}\left(\begin{array}{c}
\mathcal{K}_{1}a_{1,in}(\omega)\\
\mathcal{K}_{2}a_{2,in}(\omega)\\
\mathit{\Gamma}_{1}b_{1,in}(\omega)\\
\mathit{\Gamma}_{2}b_{2,in}(\omega)
\end{array}\right),
\end{eqnarray}
with
\begin{eqnarray*}
U_{\chi}^{-1} & = & I-\left(\begin{array}{cccc}
0 & 0 & \mathcal{A}_{11} & \mathcal{A}_{12}\\
0 & 0 & \mathcal{A}_{21}e^{i\Phi} & \mathcal{A}_{22}\\
\mathcal{B}_{11} & \mathcal{B}_{21}e^{-i\Phi} & 0 & 0\\
\mathcal{B}_{12} & \mathcal{B}_{22} & 0 & 0
\end{array}\right).
\end{eqnarray*}
This provides an intuitive way to understand the thermal noise flow.

Moreover, using the spectrum approach $s_{o}(\omega)$ $=$ $\frac{1}{2}\int_{-\infty}^{+\infty}d\omega^{\prime}\left[\langle o^{\dagger}(\omega)o(\omega^{\prime})\rangle+\langle o(\omega)o^{\dagger}(\omega^{\prime})\rangle\right]$,
we can calculate the spectrum of the cavity and mechanical modes $S(\omega)$
$=$ $[s_{a_{1}}(\omega)$, $s_{a_{2}}(\omega)$, $s_{b_{1}}(\omega)$,
$s_{b_{2}}(\omega)]^{T}$, which connect to the spectrum of the input
field $S_{in}(\omega)$ $=$ $[s_{a_{1,in}}(\omega)$, $s_{a_{2,in}}(\omega)$,
$s_{b_{1,in}}(\omega)$, $s_{b_{2,in}}(\omega)]^{T}$ by the transmission
matrix $T(\omega)$: $S(\omega)$ $=$ $T(\omega)S_{in}(\omega)$,
where the matrix elements $T_{lm}(\omega)=|U_{lm}(\omega)|^{2}$,
and the correlation functions of noises are $\langle a_{j,in}^{\dagger}(\omega)a_{j,in}(\omega^{\prime})\rangle$
$=$ $0$, $\langle a_{j,in}(\omega)a_{j,in}^{\dagger}(\omega^{\prime})\rangle$
$=$ $\delta(\omega+\omega^{\prime})$, $\langle b_{k,in}^{\dagger}(\omega)b_{k,in}(\omega^{\prime})\rangle$
$=$ $\bar{m}_{k}\delta(\omega+\omega^{\prime})$, and $\langle b_{k,in}(\omega)b_{k,in}^{\dagger}(\omega^{\prime})\rangle$
$=$ $(\bar{m}_{k}+1)\delta(\omega+\omega^{\prime})$, where $j,k=1,2$
and $\bar{m}_{k}$ $=$ $[\text{exp}(\hbar\omega_{b,k}/k_{B}T)-1]^{-1}$
is the average thermal phonon number of the $k$th mechanical mode
and $k_{B}$ is the Boltzmann constant. Then the average phonon number
are obtained by
\begin{equation}
\bar{n}_{k}=\frac{1}{2\pi}\int s_{b_{k}}(\omega)d\omega-\frac{1}{2}.\label{eq:6}
\end{equation}

For the four-mode setup, the MRs under optomechanical interactions
response to the input noises via
\begin{gather}
\overrightarrow{\chi}_{B}^{-1}(\Phi)\left(\begin{array}{c}
b_{1}\\
b_{2}
\end{array}\right)=\left[\begin{array}{cc}
1 & (\mathcal{A}_{12}\mathcal{B}_{11}+e^{-i\Phi}\mathcal{A}_{22}\mathcal{B}_{21})\chi_{\mathcal{F}}^{(12,22)}\\
(\mathcal{A}_{11}\mathcal{B}_{12}+e^{i\Phi}\mathcal{A}_{21}\mathcal{B}_{22})\chi_{\mathcal{F}}^{(11,21)} & 1
\end{array}\right]\left(\begin{array}{c}
\varGamma_{1}b_{1,in}\\
\varGamma_{2}b_{2,in}
\end{array}\right)\nonumber \\
+\left[\begin{array}{cc}
\mathcal{B}_{21}e^{-i\Phi}\mathcal{A}_{22}\mathcal{B}_{12}\chi_{\mathcal{F}}^{(12,22)}+\mathcal{B}_{11}\chi_{\mathcal{FF}}^{(12,22)} & \mathcal{B}_{11}\mathcal{A}_{12}\mathcal{B}_{22}\chi_{\mathcal{F}}^{(22,12)}+\mathcal{B}_{21}e^{-i\Phi}\chi_{\mathcal{FF}}^{(22,12)}\\
\mathcal{B}_{22}\mathcal{A}_{21}e^{i\Phi}\mathcal{B}_{11}\chi_{\mathcal{F}}^{(11,21)}+\mathcal{B}_{12}\chi_{\mathcal{FF}}^{(11,21)} & \mathcal{B}_{12}\mathcal{A}_{11}\mathcal{B}_{21}e^{-i\Phi}\chi_{\mathcal{F}}^{(21,11)}+\mathcal{B}_{22}\chi_{\mathcal{FF}}^{(21,11)}
\end{array}\right]\left(\begin{array}{c}
\mathcal{K}_{1}a_{1,in}\\
\mathcal{K}_{2}a_{2,in}
\end{array}\right),\label{eq:10-1-1-1}
\end{gather}
where $\overrightarrow{\chi}_{B}^{-1}(\Phi)=\text{diag}(D_{1},D_{2})$
and 
\begin{eqnarray*}
D_{1} & = & 1-\left(e^{-i\Phi}\mathcal{A}_{11}\mathcal{B}_{12}\mathcal{B}_{21}\mathcal{A}_{22}+e^{i\Phi}\mathcal{B}_{11}\mathcal{A}_{12}\mathcal{A}_{21}\mathcal{B}_{22}\right)\chi_{\mathcal{F}}^{(12,22)}\\
 &  & -\mathcal{A}_{11}\mathcal{B}_{11}\chi_{\mathcal{FF}}^{(12,22)}-\mathcal{A}_{21}\mathcal{B}_{21}\chi_{\mathcal{FF}}^{(22,12)},
\end{eqnarray*}
\begin{eqnarray*}
D_{2} & = & 1-\left(e^{-i\Phi}\mathcal{A}_{11}\mathcal{B}_{12}\mathcal{B}_{21}\mathcal{A}_{22}+e^{i\Phi}\mathcal{B}_{11}\mathcal{A}_{12}\mathcal{A}_{21}\mathcal{B}_{22}\right)\chi_{\mathcal{F}}^{(11,21)}\\
 &  & -\mathcal{A}_{12}\mathcal{B}_{12}\chi_{\mathcal{FF}}^{(11,21)}-\mathcal{A}_{22}\mathcal{B}_{22}\chi_{\mathcal{FF}}^{(21,11)}.
\end{eqnarray*}

\section*{Appendix C thermal noise controlled by the driving laser}

\begin{figure}[H]
\begin{centering}
\includegraphics[width=0.6\columnwidth]{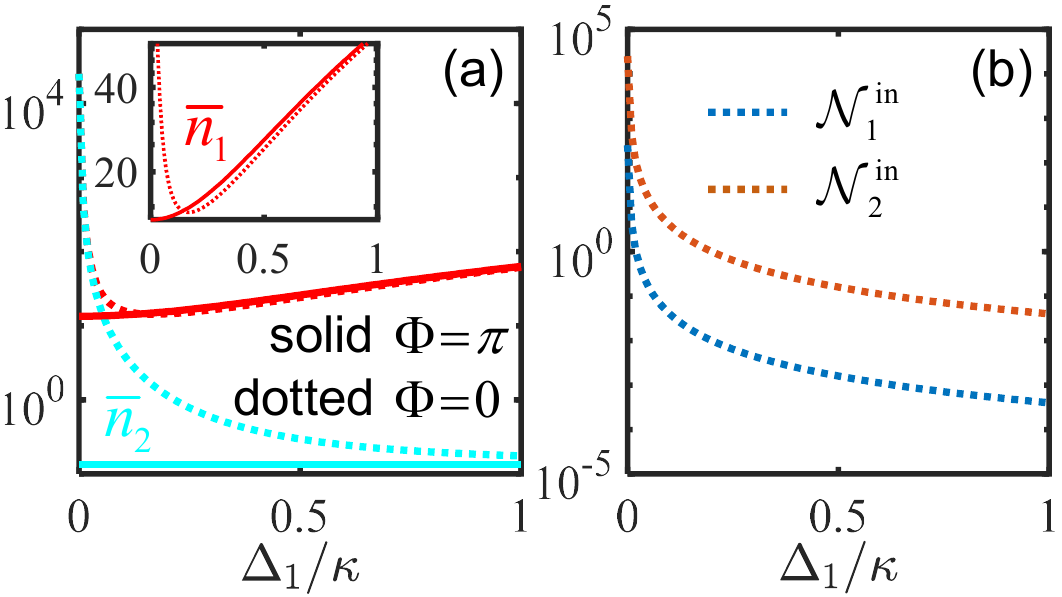}
\par\end{centering}
\caption{\label{fig:n_delta}(a) Phonon occupation numbers $\bar{n}_{1}$ (red
curves) and $\bar{n}_{2}$ (cyan curves) as functions of $\Delta_{1}/\kappa$
for $\Phi=0$ (dotted curves) and $\Phi=\pi$ (solid curves). (b)
$\mathcal{N}_{1}^{\text{in}}$ (blue dotted curve) and $\mathcal{N}_{2}^{\text{in}}$
(orange dotted curve) as functions of $\Delta_{1}/\kappa$ with $\Delta_{2}=0$,
where the overall plaquette phase is $\Phi=0$. The thermal phonon
number for $R_{1}$ and $R_{2}$ considered here are $\bar{m}_{1}=10^{5}$
and $\bar{m}_{2}=10^{3}$. Other parameters are the same as in Fig.
\ref{fig:model}(b).}
\end{figure}

For $\Phi=0$, the two MRs can not be cooled efficiently by the cold
cavity reservoir. However, by introducing appropriately imbalanced
laser detunings, e.g. $\Delta_{1}$ $\neq$ $0$ and $\Delta_{2}=0$,
MR2 can be optomechanically cooled, as shown in Fig. \ref{fig:n_delta}(a).
Although the noise flow between two MRs can be reduced as the detuning
$\Delta_{1}$ increases, see Fig. \ref{fig:n_delta}(b), only MR2
can approach the dual-cavity cooling limit as if MR1 could be adiabatically
eliminated. Cooling MR1 is significantly improved with an extra small
detuning, but becomes inefficient due to the unresonant sideband driving
when the detuning further increases. While for $\Phi=\pi$, because
the noise flow between two MRs is eliminated completely, i.e. $\mathcal{N}_{1}^{\text{in}}$
$=$ $\mathcal{N}_{2}^{\text{in}}$ $=0$, the cooling of MR2 reaches
the dual-cavity cooling limit and is not affected by $\Delta_{1}/\kappa$,
but again, the optomechanical cooling of MR1 becomes inefficient as
$\Delta_{1}/\kappa$ increases.

In Fig. \ref{fig:n_G21}, we show the phonon occupation number of
the MR2 $\bar{n}_{2}$ as a function of $G_{21}/G_{12}$ for $\bar{m}_{1}=$
$\{10^{3},10^{4},5\times10^{4},10^{5}\}$. Obviously, the dual-cavity
cooling limit of MR2 is obtained by decoupling from MR1 (i.e. $G_{11}$
$=$ $G_{21}$ $=$ $0$). When the impedance condition $G_{21}/G_{12}$
$=(G_{22}/G_{11})^{-1}$ is not fulfilled, the mechanical cooling
becomes far from the dual-cavity cooling limit because of the influence
of the thermal noise $\bar{m}_{1}$. An increasing $\bar{m}_{1}$
further reduces the cooling efficiency of MR2. For comparison, we
show that ground-state cooling cannot be achieved for $\bar{m}_{1}>5\times10^{4}$
under the dark-mode breaking regime $G_{21}/G_{12}=G_{22}/G_{11}$,
as indicated by the red vertical line. Remarkably, we emphasize that
under the fulfillment of $G_{21}/G_{12}$ $=(G_{22}/G_{11})^{-1}$,
the cooling effect of MR2 precisely matches the dual-cavity cooling
limit and is independent of $\bar{m}_{1}$. While for MR1 itself,
which is not shown in the figure, the cooling efficiency (under fixed
coupling strengths) drops significantly due to the increased thermal
temperature.
\begin{figure}[H]
\centering{}\includegraphics[width=0.45\columnwidth]{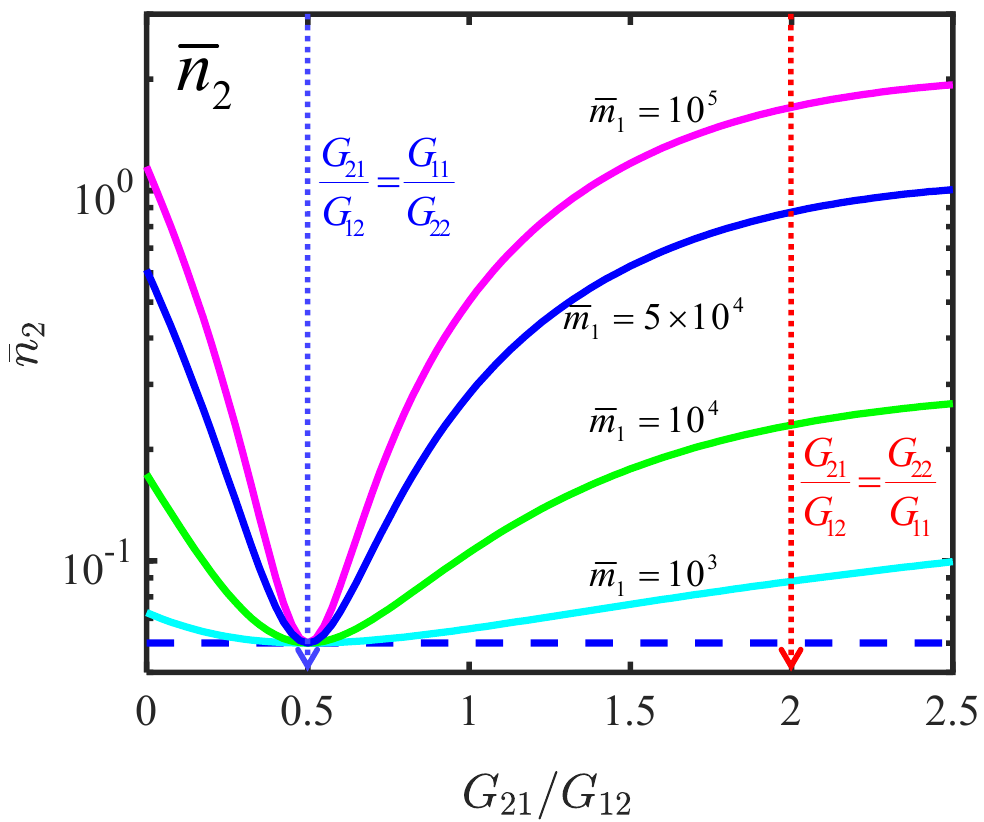}\caption{\label{fig:n_G21}Phonon occupation number $\bar{n}_{2}$ as a function
of $G_{21}/G_{12}$ with $G_{22}/G_{11}=2$ for a set of thermal phonon
number $\bar{m}_{1}=$ $\{10^{3},10^{4},5\times10^{4},10^{5}\}$ and
$\bar{m}_{2}=10^{3}$. The horizontal dashed curve corresponds to
the dual-cavity cooling limit of MR2 in the $a_{1}-b_{2}-a_{2}$ setup.
The vertical dotted line on the left (right) indicates the impedance
condition (the conventional dark-mode breaking regime). Other parameters
are the same as in Fig. \ref{fig:n_G22_G21}.}
\end{figure}

\section*{Appendix D nonreciprocal phonon transfer}

\begin{figure}[H]
\centering{}\includegraphics[width=0.7\columnwidth]{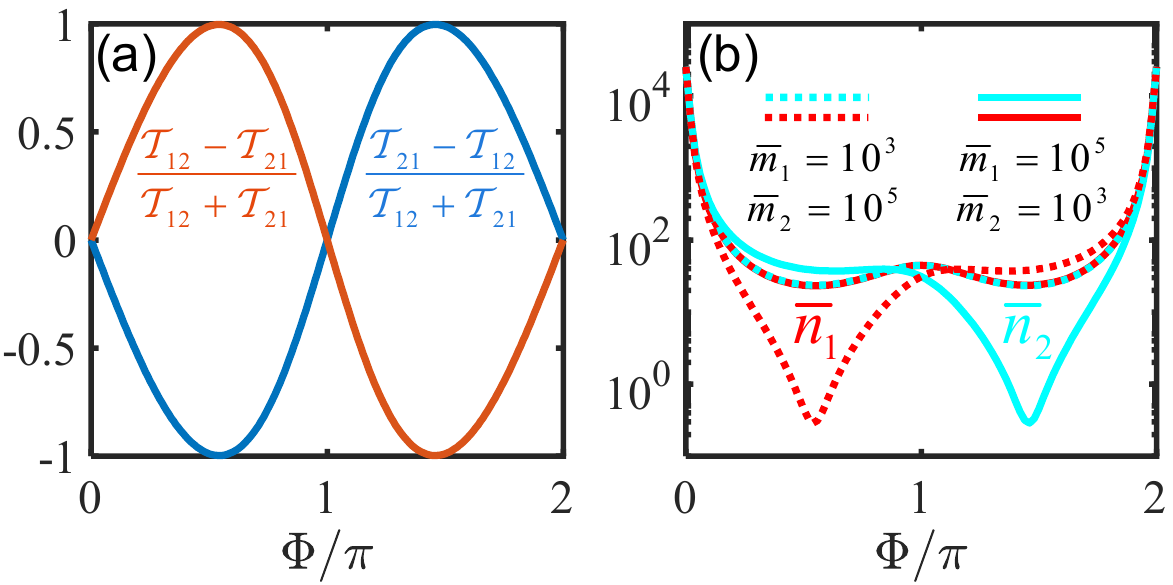}\caption{\label{fig:nonreciprocal}(a) The transmission between the two MRs,
$(\mathcal{T}_{21}-\mathcal{T}_{12})/(\mathcal{T}_{12}+\mathcal{T}_{21})$
(blue solid curve) and $(\mathcal{T}_{12}-\mathcal{T}_{21})/(\mathcal{T}_{12}+\mathcal{T}_{21})$
(orange solid curve) as functions of $\Phi/\pi$, (b) Phonon occupation
numbers $\bar{n}_{1}$ (red curves) and $\bar{n}_{2}$ (cyan curves)
as functions of $\Phi/\pi$ at the cases of $\bar{m}_{1}=10^{5}$,
$\bar{m}_{2}=10^{3}$ (solid curves) and $\bar{m}_{1}=10^{3}$, $\bar{m}_{2}=10^{5}$
(dotted curves). Parameters used here are $\tilde{\Delta}_{1}/\kappa=0$,
$\tilde{\Delta}_{2}/\kappa=0.5$, and $\kappa_{2}/\kappa=0.1$. Other
parameters are the same as in Fig. \ref{fig:model}(b).}
\end{figure}

So far the thermal phonon flow between the two MRs is reciprocal under
the Hamiltonian of Eq. (\ref{eq:H_1}), i.e. $T_{R_{2}\rightarrow b_{1}}=T_{R_{1}\rightarrow b_{2}}$.
But it can be turned into nonreciprocal (and even uni-directional)
\citep{Liang_2010,Boechler_2011,Popa_2014,Fleury2014,Xu_2016,Sasaki2017,Xu2019,Chen2021a,Lan2022,Shen2023,Tang2023}
by considering the cavity driving frequencies $\omega_{a,j}-\omega_{b,k}+\tilde{\Delta}_{j}$,
which leads to the alternative linearized Hamiltonian 
\begin{eqnarray}
H_{2} & = & \sum_{j=1,2}\tilde{\Delta}_{j}a_{j}^{\dagger}a_{j}+(G_{11}a_{1}b_{1}^{\dagger}+G_{12}a_{1}b_{2}^{\dagger}\nonumber \\
 &  & +G_{21}e^{-i\Phi}a_{2}b_{1}^{\dagger}+G_{22}a_{2}b_{2}^{\dagger}+\text{H.c.}).
\end{eqnarray}
The coupled equation for MRs is again given by Eq. (\ref{eq:NoiseFlow_4ModeA}),
with the response functions being revised as $\chi_{a,j}$ $=$ $(\kappa_{j}/2+i\tilde{\Delta}_{j}-i\omega)^{-1}$
and $\chi_{b,k}$ $=$ $(\gamma_{k}/2-i\omega)^{-1}$ ($j,k$ $=$
$1,2$). Furthermore, by considering $\tilde{\Delta}_{2}$ $\gg$
$\{\kappa_{2},G_{21},G_{22}\}$ (or $\kappa_{1}$ $\gg$ $\{\tilde{\Delta}_{1},G_{11},G_{12}\}$)
one can adiabatically eliminate the optical degree of freedom for
the coherent path $b_{1}$ $\rightarrow$ $a_{1}$ $\rightarrow$
$b_{2}$ (the dissipative path $b_{1}$ $\rightarrow$ $a_{2}$ $\rightarrow$
$b_{2}$), where the matrix elements $\mathcal{H}_{12}=(\mathcal{A}_{12}\mathcal{B}_{11}+e^{-i\Phi}\mathcal{A}_{22}\mathcal{B}_{21})\chi_{\mathcal{F}}^{(12,22)}$
and $\mathcal{H}_{21}=(\mathcal{A}_{11}\mathcal{B}_{12}+e^{i\Phi}\mathcal{A}_{21}\mathcal{B}_{22})\chi_{\mathcal{F}}^{(11,21)}$
in Eq. (\ref{eq:NoiseFlow_4ModeA}) vanish if 
\[
\left|\frac{G_{11}G_{12}}{\kappa_{1}/2}+e^{-i\Phi}\frac{G_{21}G_{22}}{i\tilde{\Delta}_{2}}\right|=0,
\]
\begin{equation}
\left|\frac{G_{11}G_{12}}{\kappa_{1}/2}+e^{i\Phi}\frac{G_{21}G_{22}}{i\tilde{\Delta}_{2}}\right|=0.
\end{equation}
Then, a uni-directional transfer of thermal phonons can be realized
with $\mathcal{H}_{12}$ $=$ $0$ (or $\mathcal{H}_{21}$ $=$ $0$),
i.e. $\mathcal{T}_{21}=0$ ($\mathcal{T}_{12}=0$) for $\Phi=\pi/2$
($\Phi=3\pi/2$), as shown in Fig. \ref{fig:nonreciprocal}(a). Then,
the ground-state cooling of MR2 with $\bar{n}_{2(1)}$ $=0.23$ is
achieved for $\bar{m}_{1}=10^{5}$ and $\bar{m}_{2}=10^{3}$ ($\bar{m}_{1}=10^{3}$
and $\bar{m}_{2}=10^{5}$) at around $\Phi=3\pi/2$ ($\Phi=\pi/2$)
since the thermal noise flow from $R_{1(2)}$ is inhibited, but MR1(2)
stays far away its quantum ground state due to a large $\bar{m}_{1(2)}$
{[}see Fig. \ref{fig:nonreciprocal}(b){]}. Interestingly, one can
then realize phonon routing with tens of thermal phonons flowing between
the two MRs.

\end{widetext}
\end{document}